\documentclass[sigconf]{acmart}
\usepackage{amsmath,amsfonts}
\usepackage{algorithmic}
\usepackage{graphicx}
\usepackage{xcolor}
\usepackage{hyperref}
\usepackage{setspace}
\usepackage{booktabs} 
\usepackage{subfigure}
\usepackage{amsmath}
\usepackage{color}
\usepackage{amsmath}
\usepackage{mathrsfs}
\usepackage[normalem]{ulem}
\usepackage{enumitem}
\usepackage{multirow}
\usepackage{ulem}
\usepackage[nomargin,inline,marginclue,draft]{fixme}
\usepackage{balance}
\usepackage{verbatim}
\usepackage{diagbox}
\usepackage{changepage}
\usepackage{xcolor}
\usepackage{bm}
\usepackage{booktabs}
\usepackage{multirow}
\usepackage{makecell}
\usepackage{bbding}
\usepackage{balance}

\AtBeginDocument{%
  }

\copyrightyear{2023}
\acmYear{2023}
\setcopyright{acmlicensed}\acmConference[KDD '23]{Proceedings of the 29th ACM SIGKDD Conference on Knowledge Discovery and Data Mining}{August 6--10, 2023}{Long Beach, CA, USA}
\acmBooktitle{Proceedings of the 29th ACM SIGKDD Conference on Knowledge Discovery and Data Mining (KDD '23), August 6--10, 2023, Long Beach, CA, USA}
\acmPrice{15.00}
\acmDOI{10.1145/3580305.3599922}
\acmISBN{979-8-4007-0103-0/23/08}

\begin{document}

\title{TWIN: TWo-stage Interest Network for Lifelong \\User Behavior Modeling in CTR Prediction at Kuaishou}

\author{Jianxin Chang}
\affiliation{
  \institution{Kuaishou Technology}
  \city{Beijing}
  \country{China}}
\email{changjianxin@kuaishou.com}

\author{Chenbin Zhang}
\affiliation{
  \institution{Kuaishou Technology}
  \city{Beijing}
  \country{China}}
\email{zhangchenbin@kuaishou.com}

\author{Zhiyi Fu}
\affiliation{
  \institution{Kuaishou Technology}
  \city{Beijing}
  \country{China}}
\email{fuzhiyi@kuaishou.com}

\author{Xiaoxue Zang}
\affiliation{
  \institution{Kuaishou Technology}
  \city{Beijing}
  \country{China}}
\email{zangxiaoxue@kuaishou.com}

\author{Lin Guan}
\affiliation{
  \institution{Kuaishou Technology}
  \city{Beijing}
  \country{China}}
\email{guanlin03@kuaishou.com}

\author{Jing Lu}
\authornote{Corresponding Author}
\affiliation{
  \institution{Kuaishou Technology}
  \city{Beijing}
  \country{China}}
\email{lvjing06@kuaishou.com}

\author{Yiqun Hui}
\affiliation{
  \institution{Kuaishou Technology}
  \city{Beijing}
  \country{China}}
\email{huiyiqun@kuaishou.com}

\author{Dewei Leng}
\affiliation{
  \institution{Kuaishou Technology}
  \city{Beijing}
  \country{China}}
\email{lengdewei@kuaishou.com}

\author{Yanan Niu}
\affiliation{
  \institution{Kuaishou Technology}
  \city{Beijing}
  \country{China}}
\email{niuyanan@kuaishou.com}

\author{Yang Song}
\affiliation{
  \institution{Kuaishou Technology}
  \city{Beijing}
  \country{China}}
\email{yangsong@kuaishou.com}

\author{Kun Gai}
\affiliation{
  \institution{Unaffiliated}
  \city{Beijing}
  \country{China}}
\email{gai.kun@qq.com}

\renewcommand{\shortauthors}{Jianxin Chang et al.}

\begin{abstract}
Life-long user behavior modeling, i.e., extracting a user's hidden interests from rich historical behaviors in months or even years, plays a central role in modern CTR prediction systems. 
Conventional algorithms mostly follow two cascading stages: a simple General Search Unit (GSU) for fast and coarse search over tens of thousands of long-term behaviors and an Exact Search Unit (ESU) for effective Target Attention (TA) over the small number of finalists from GSU. 
Although efficient, existing algorithms mostly suffer from a crucial limitation: the \textit{inconsistent} target-behavior relevance metrics between GSU and ESU. As a result, their GSU usually misses highly relevant behaviors but retrieves ones considered irrelevant by ESU. In such case, the TA in ESU, no matter how attention is allocated, mostly deviates from the real user interests and thus degrades the overall CTR prediction accuracy. 
To address such inconsistency, we propose \textbf{TWo-stage Interest Network (TWIN)}, where our Consistency-Preserved GSU (CP-GSU) adopts the identical target-behavior relevance metric as the TA in ESU, making the two stages twins. Specifically, 
to break TA's computational bottleneck and extend it from ESU to GSU, or namely from behavior length $10^2$ to length $10^4-10^5$, we build a novel attention mechanism by behavior feature splitting.
For the video inherent features of a behavior, we calculate their linear projection by efficient pre-computing \& caching strategies.
And for the user-item cross features, we compress each into a one-dimentional bias term in the attention score calculation to save the computational cost.
The consistency between two stages, together with the 
effective TA-based relevance metric in CP-GSU, contributes to significant performance gain in CTR prediction. Offline experiments on a 46 billion scale real production dataset from Kuaishou and an Online A/B test show that TWIN outperforms all compared SOTA algorithms.
With optimized online infrastructure, 
we reduce the computational bottleneck by 99.3\%, which contributes to the successful deployment of TWIN on Kuaishou, serving the main traffic of hundreds of millions of active users everyday.

\end{abstract}




\maketitle

\section{Introduction} \label{sec:intro}
As one of the most popular short video sharing apps in China, \textit{Kuaishou} strongly relies on its powerful recommendation system (RS). 
Every day, RS helps hundreds of millions of active users to filter out millions of uninterested videos and reach their interested ones,
leaving tens of billions of click through logs.
These tremendous data not only feed the training of RS, but also boost technique revolutions that keep lifting both the user experience and business effectiveness on this platform.

In modern RSs, a fundamental task is Click Through Rate (CTR) prediction which aims to 
predict the probability that a user would click an item / video \cite{chen2016deep,guo2017deepfm,zhang2019field}. Accurate CTR prediction directs RS to serve each user one's favourite contents and deliver each video to its interested audiences. To achieve this, CTR models should be highly personalized and make full use of  
scarce user information.
Consequently, \textit{life-long user behavior modeling}, i.e., extracting a user's hidden interests from rich long-term historical behaviors, usually acts as a key component in CTR models %and attracts widespread interest in both machine learning academia and RS industry 
\cite{zhou2018deep,zhou2018atrank,liu2007framework,zhou2019deep,chung2014empirical}.

Industrial life-long behavior modeling algorithms mostly follow the two cascading stages \cite{pi2020search}: 
(1) a General Search Unit (GSU) that conducts a fast coarse search 
over tens of thousands of long-term behaviors and outputs a small number of most target-relevant ones, and 
(2) an Exact Search Unit (ESU) that
performs effective 
\textit{Target Attention} (TA) over the small number of finalists from GSU.
The reason behind this two-stage design is twofold. On one hand, to precisely capture the user interest, TA is a proper choice
for emphasizing target-relevant behaviors and suppressing target-irrelevant ones. On the other hand, the 
expensive computational cost of TA limits its applicable sequence length to at most a few hundreds. 
To this end, a simple and fast GSU as a pre-filter is essential for cutting-off industrial scale behavior sequences which could easily reach $10^4$-$10^5$ in just a few months.

\begin{figure}[t]
  \centering
  \hspace{-0.3cm}
  \includegraphics[width=0.38\textwidth]{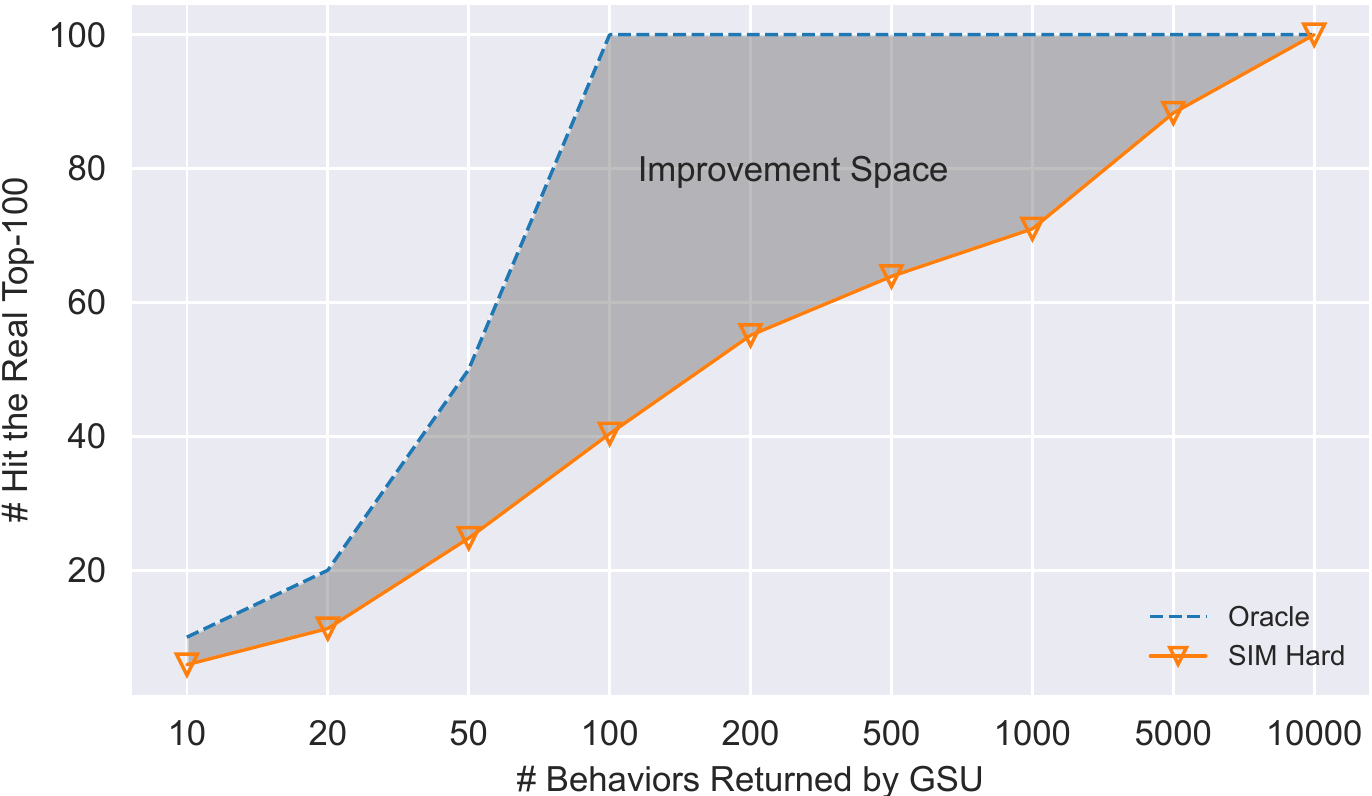}
  \caption{Inconsistency between GSU \& ESU in a Conventional Two-Stage Algorithm. 
  Assume that an 
  ``Oracle'' (blue) could afford to use the identical relevance metric as that in ESU on all the $10^4$-$10^5$ behaviors, namely find ``the real top-100''.
While GSU (orange) uses an ineffective and inconsistent coarse search. Among the top-100 returned by GSU 
  (x axis), only 40 hit the real top-100 (y axis). This inconsistency (gray) indicates the potential improvement space left for TWIN.  
  } 
  \vspace{-0.4cm}
  \label{fig:inconsist}
\end{figure}

Recent years have witnessed a great many emerging studies on two-stage life-long behavior modeling, while their key difference lies in the GSU strategies that coarsely select target-relevant behaviors.
For example, SIM Hard \cite{pi2020search} simply selects behaviors from the same category as the target item, 
while SIM Soft \cite{pi2020search} calculates the target-behavior relevance score from pre-trained item embeddings by the inner product and selects behaviors with the highest relevance \cite{pi2020search}.
ETA approximates the relevance score calculation using locality-sensitive hashing (LSH) and Hamming distance \cite{chen2021end}. 
SDIM samples behaviors with the same hash signature as the target-behavior through multi-round hash collision \cite{cao2022sampling}, among others.
Despite being extensively studied, existing two-stage life-long behavior modeling algorithms still suffer from a crucial limitation:
the \textit{inconsistency} between GSU and ESU (shown in Figure \ref{fig:inconsist}\footnote{The Oracle is identically copied from the ESU of SIM-hard (trained from the top-100 behaviors retrieved by GSU), 
except that it ranks $10^4$ rather than $10^2$ behaviors.
Experiments are conducted on a tiny demo dataset from Kuaishou, not feasible for normal datasets due to the extremely high cost of Oracle (details in Section \ref{sec:rq2}). 
}). Specifically, the target-behavior relevance metric used in GSU is both coarse and inconsistent with the TA used in ESU. 
As a result, GSU may probably miss relevant behaviors, 
but retrieve ones considered irrelevant by ESU, wasting ESU's precious computational resources.
In such case, the TA in ESU, no matter how attention is allocated, mostly deviates from the real user interests and thus degrades the overall CTR prediction accuracy.

To address such \textit{inconsistency}, we propose TWIN: TWo-stage Interest Network for lifelong user behavior modeling, where 
a Consistency-Preserved GSU (CP-GSU) adopts the \textit{identical} target-behavior relevance metric as the TA in ESU, making the two stages \textit{twins}. 
To extend the expensive TA to CP-GSU,
TWIN breaks TA's key computational bottleneck, namely the linear projection of all behaviors, by effective behavior feature split, simplified TA architecture and highly optimized online infrastructure.
1). Specifically, for the video inherent features of a behavior (e.g. video id, author, duration, topic) which are shared across users / behavior sequences,
we accelerate their projection by 
efficient pre-computing \& caching strategies.
2). And for the the user-video cross features of a behavior (e.g. user's click timestamp, play time, rating), where caching is not applicable, we 
simplify the TA architecture by compressing their projection into bias terms. 
With optimized online infrastructure, we successfully extend the applicable sequence length of TA from $10^2$ in ESU to $10^4-10^5$ in CP-GSU.
The consistency between two stages, together with the 
effective 
TA-based relevance metric in CP-GSU, contributes to significant performance gain in CTR prediction.

Overall, we make the following contributions: %\vspace{-0.1cm}
\begin{itemize}[leftmargin=*]
    \item In our proposed TWIN, CP-GSU precisely and consistently retrieves behaviors that are not only target-relevant, but also considered important by ESU, maximizing the retrieval effectiveness of behavior modeling.
    To the best of our knowledge, we are the first to successfully address the inconsistency in the two-stage life-long behavior modeling problem. 
    
    \item We validate the effectiveness of TWIN through extensive offline experiments on Kuaishou's 46 billion scale industrial dataset and online A/B tests.
     We verify our validity through ablation studies and show that TWIN brings significant online benefits.
    
    \item We build efficient industrial infrastructure to apply TWIN on the real online RS. Effective pre-computing \& caching strategies are proposed to reduce the computational bottleneck of TWIN, i.e., the linear projection of behaviors in CP-GSU, by 99.3\% and to meet the low latency requirements of the online serving system. TWIN has now been deployed on the RS of Kuaishou, serving the main traffic of 346 million active users every day.
\end{itemize}

\section{Related Work} \label{sec:related_work}
Our work is closely related to two active research areas: CTR prediction and long-term user behavior modeling.
\vspace{-2px}
\subsection{Click-Through-Rate Prediction}
CTR prediction which aims to predict a user's personalized interests, is crucial for nowadays RSs. Early CTR models are shallow and mainly focus on exploiting feature interactions, such as factorization machines (FM) \cite{rendle2010factorization} and field-aware factorization machines (FFM) \cite{juan2016field}. With the success of deep learning, deep CTR models are extensively studied and become mainstream choices. For example, \citet{chen2016deep} and \citet{zhang2016deep} first apply deep models for CTR tasks. Wide\&Deep \cite{cheng2016wide} combines a wide linear model and a deep model, which takes the advantages of both memorization of feature interaction and generalization of deep architecture. DeepFM \cite{guo2017deepfm} and DCN \cite{wang2017deep, wang2021dcn} improve the wide part of Wide\&Deep to increase the feature interaction ability. xDeepFM \cite{lian2018xdeepfm} and AFM \cite{xiao2017attentional} further exploit convolution-like layers and attention mechanism to improve the deep part and boost model performance.

As the CTR models becomes increasingly personalized, user behavior modeling, i.e., 
capturing a user's hidden interest from the summarization of one's historical behaviors,
becomes a crucial module.
Limited by computational resources, early algorithms are mostly in target-independent manners and thus can be efficiently pre-calculated offline \cite{covington2016deep,yu2016dynamic,song2016multi}. To better extract a user's interest in specific items, various TA mechanisms are adopted.
DIN \cite{zhou2018deep}, 
represents the user interest by a TA over historical behaviors to emphasise target-relevant behaviors.
DIEN \cite{zhou2019deep} further introduces the temporal relationship of behaviors using ARGRU, an attention-based variant of classic GRU \cite{cho2014learning}. DSIN \cite{feng2019deep} splits behaviors into multiple sessions and conducts self-attention inside each one to
emphasise intra-session relationships. MIND \cite{li2019multi} and DMIN \cite{xiao2020deep} 
represents user interest by multiple vectors. 
BST \cite{chen2019behavior}, SASRec \cite{kang2018self} and BERT4Rec \cite{sun2019bert4rec} also use transformers to improve the model's performance and parallelism.

\begin{table}[t]
\small
\caption{Comparison of SOTA user interest models. The bottom part lists the two-stage models. Length denotes the maximum sequence length of user behaviors in original papers.}
\begin{tabular}{lllcc}
\toprule
\multirowcell{2}{Method} & \multirowcell{2}{Length} & \multirowcell{2}{GSU Stragegy} & \multirowcell{2}{End2End} &\multirowcell{2}{Consist}  \\
\\
\midrule
DIN \cite{zhou2018deep} & $\sim 10^3$ & N/A & N/A & N/A\\
DIEN \cite{zhou2019deep} & $\sim 10^2$ & N/A & N/A & N/A\\
MIMN \cite{pi2019practice} & $\sim 10^3$ & N/A & N/A & N/A\\
\midrule
UBR4CTR \cite{qin2020user,qin2023learning} & $\sim 10^2$ & BM25 & \XSolidBrush& \XSolidBrush \\
SIM Hard \cite{pi2020search} & $\sim 10^3$ & Category Filter & \XSolidBrush& \XSolidBrush \\
SIM Soft \cite{pi2020search} & $\sim 10^3$ & Inner Product & \XSolidBrush& \XSolidBrush \\
{ETA \cite{chen2021end}} &{$\sim 10^3$} & LSH \& Hamming  & {\Checkmark}& {\XSolidBrush}\\
{SDIM \cite{cao2022sampling}} &{$\sim 10^3$} &Hash Collision & {\Checkmark}&{\XSolidBrush} \\
TWIN (ours) & $\sim 10^5$ & Target Attention & \Checkmark & \Checkmark\\
\bottomrule
\label{tab:comparison}
\end{tabular}
\vspace{-20px}
\end{table}

\begin{table}
\caption{Important Notations Used in Section 3}\label{notations}
\begin{center}
\begin{tabular}{ll|ll|ll}
   \toprule
		  $f$ & predictor		&	$\sigma$ &sigmoid 			&$\hat y$ & predicted CTR	\\
$\mathcal D$&dateset			&	$\mathbb R$ & real number set  	&$d$& feature dimension\\
$\ell$&loss&$\mathbf x$&feature vector		&	$y$& ground truth label\\
\midrule
\end{tabular}
\begin{tabular}{ll|ll}
 $E$ & embedding dictionary&$\mathbf x_\text{emb}$ &embedded feature \\
$\mathbf x_\text{hot}$& one/multi hot coding &$v$& vocabulary size	\\
$K$ & behavior features & $L$ & behavior length \\
$J$ & cross feature number &$a$ & head index  \\
$H$ & inherent feature dimension & $K_h$ & inherent features \\
$C$ & cross feature dimension &  $K_c$ & cross features\\ 
$\boldsymbol \beta$ & cross feature weight &$\boldsymbol \alpha$ & attention weight \\
\midrule
\end{tabular}\end{center}
\begin{tabular}{ll}
$d_k$ & dimension of projected inherent features  \\
$d_{out}$ & dimension of projection in original MHTA \\
$W^h, W^c, W^v, W^o$ & linear projection parameters \\
$\mathbf w_j^c$ & a diagonal block of projection parameter $W^c$ \\
\bottomrule
\end{tabular}
\vspace{-10px}
\end{table}

\vspace{-3px}
\subsection{Long-Term User Behavior Modeling}
As the effectiveness of TA and interest modeling have been confirmed in modern industrial RSs, researchers start to model increasingly longer behaviors. % and try to dig out fine-grained interests from different user behavior stages. For example, 
\citet{liu2007framework} combines long-term and short-term interests in CTR prediction. MIMN \cite{pi2019practice} stores the user behaviors as a memory matrix at the user interest center (UIC) and updates the memory when new user behavior comes. However, MIMN is  hard to be extended to sequences longer than $10^3$, and it generates the same memory matrix for different candidate items, carrying useless noise and impairing the TA. 

Recently, SIM \cite{pi2020search} and UBR4CTR \cite{qin2020user,qin2023learning} introduce the two-stage cascaded framework to solve those challenges and achieve SOTA performance in CTR prediction. A conventional two-stage algorithm usually consists of:
1) a simple and fast GSU which retrieves the most ``relevant'' items to the target item from thousands of user behaviors, 2). and an attentive ESU to perform TA over the finalists from GSU. 
UBR4CTR utilizes BM25
as the relevance metric in its first stage.
While in the original SIM, there are two instances with different GSU designs. The GSU of SIM Hard
selects relevant items from the same category with the target item, while the GSU of SIM Soft uses the inner product of pre-trained item embeddings as the 
relevance metric. 
Though the two-stage design takes a big step forward, the original GSU still faces high computational burden and has different retrieval metric with ESU, which results in the inconsistency between two stages. 

More recently, ETA \cite{chen2021end} uses locality-sensitive hash (LSH) to encode item embeddings trained by ESU and retrieves relevant items from long-term behaviors via Hamming distance (HD). SDIM \cite{cao2022sampling} samples behavior items with the same hash signature as the target item through multiround hash collision, and the ESU then linearly aggregates those sampled behavior items to obtain user interests.
It is positive that ETA and SDIM adopt End2End training. In  other words, their 
two stages share the identical embeddings.
Nevertheless, inconsistency still exists in the retrieval strategies, specifically the network structure and parameters. 

\begin{figure*}[t]
  \centering
  \hspace{-0.3cm}
  \includegraphics[width=0.93\textwidth]{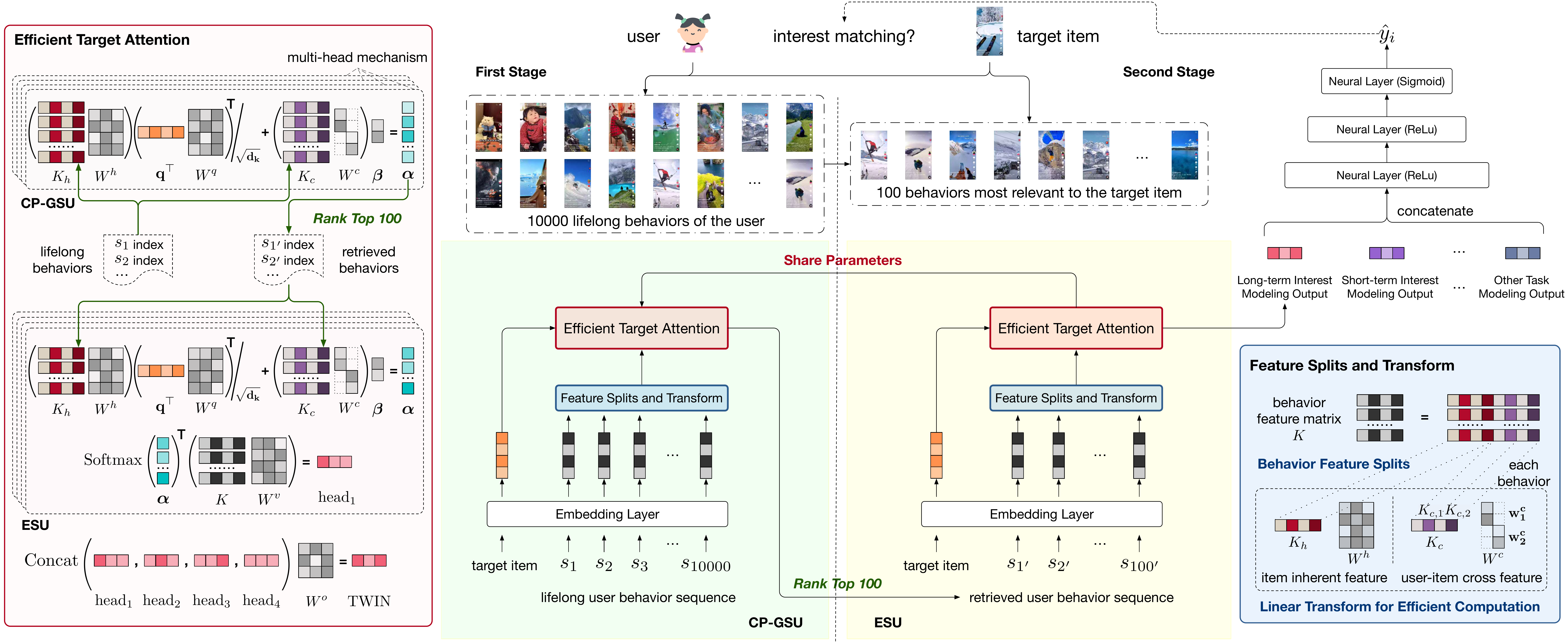}
  \caption{TWIN in Kuaishou's CTR prediction system.
   Different from conventional two-stage behavior modeling algorithms,
   TWIN adopts identical target-behavior relevance metric in CP-GSU and ESU, including not only identical network architecture (shown in the left) but also identical parameter values (shown in the middle bottom). This is challenging since MHTA was designed with high computational cost and thus only applicable to ESU (with 100 behaviors), not to CP-GSU (with $10^4$ behaviors).
    We address this challenge by proposing: 1). efficient feature split and projection strategies that process item inherent features and user-item cross features in different manners (shown in the right bottom); 2). simplified target attention architecture that accelerates the efficiency of target attention through compressing cross features into bias terms
    (shown in the left). 
}
  \vspace{-0.2cm}
  \label{fig:framework}
\end{figure*}

In this paper, we propose to extend the TA structure to GSU and synchronize the embeddings and attention parameters from ESU to GSU, maintaining the end-to-end training. As the result, we achieve the consistency in both network structure and model parameters, which contributes to significant performance gain compared to ETA and SDIM.
We detail the differences of our model with others in Table \ref{tab:comparison}.
Note that our work differs from indexing algorithms that aim to speed up transformer (e.g., LISA\cite{wu2021linear}). They approximate the relevance score calculation by mapping behaviors to codebooks and looking up the distance. While our work, as well as many other two-stage algorithms, uses exact distance calculation but reduces the number of behaviors using GSU as the pre-filter.

\section{TWIN in KuaiShou CTR Prediction} \label{sec:method}
At the beginning, we first review the general preliminaries of the CTR prediction problem 
in Section \ref{sec:preliminaries}. Then we describe the model architecture of our CTR prediction system in Kuaishou in Section \ref{sec: architecture}. We further dig into details of our proposed  
 \textit{consistency-preserved}
lifelong user behavior modeling module, named
TWo-stage Interest Network (TWIN), 
in Section \ref{sec: TWIN}. Finally, we introduce essential accelerating strategies which guarantee the successful online deployment of TWIN on the main traffic of Kuaishou in Section \ref{sec:deploy}. The notations used are summarized in Table \ref{notations}. 

\vspace{-5px}
\subsection{Preliminaries}\label{sec:preliminaries}
The aim of CTR prediction is to predict the probability that a \textit{user} would $click$ an \textit{item} given specific \textit{contexts}. 
Accurate CTR prediction not only lifts the user experience by serving preferred contents, but also benefits the business effectiveness of content producers and platforms by reaching interested audiences. 
Consequently, CTR prediction has become the core component in various industrial RSs, especially short video recommendation platforms like Kuaishou.

CTR prediction is usually formulated as a binary classification problem, where the goal is to learn a predictor function $f:\mathbb R^d \rightarrow \mathbb R$ given a training dataset $\mathcal D =\{(\mathbf x_1,y_1),...,(\mathbf x_{|\mathcal D|}, y_{|\mathcal D|})\}$. 
Specifically, $\mathbf{\mathbf x_i\in\mathbb R^d}$ is the feature vector of the $i$-th training sample (namely, the concatenation of the user, item and contexts features), and $y_i \in \{0,1\}$ is the ground truth label denoting whether the user clicks (1) the item or not (0).
The predicted CTR is calculated as:
\begin{equation}
\hat y_i = \sigma(f(\mathbf x_i)).
\end{equation}
$\sigma(\cdot)$ is the sigmoid function that scales the prediction of $f$ to $(0,1)$.

The model is trained by minimizing the negative log-likelihood:
\begin{equation}
    \ell(\mathcal D) = -\frac{1}{|\mathcal D|} \sum_{i=1}^{|\mathcal D|} y_i \log(\hat y_i) + (1-y_i)\log(1-\hat y_i).
\end{equation}
For conciseness, we omit the training sample index $i$ in the following sections when no confusion is caused.

\subsection{The Architecture of the CTR Prediction}\label{sec: architecture}
We now illustrate the architecture of our CTR prediction system at Kuaishou. Details are shown in Figure \ref{fig:framework}.
\subsubsection{The Embedding Layer}
At the bottom, out model starts from a feature embedding layer that transforms raw features of a training sample to embedding vectors.

Without loss of generality, we assume that all features are in the categorical form after essential pre-processing. For a feature $A$ with vocabulary size $v_A$, we first encode the categorical information into a one-hot / multi-hot code $\mathbf x_{\text{A},\text{hot}} \in \{0,1\}^{v_A}$. For example,

\textsf{WeekDay=Mon} ~~ ~~~~~$\Longrightarrow$ $\mathbf x_\text{WeekDay, hot} = [1,0,0,0,0,0,0]^\top$,

 \textsf{Topic=\{Funny, Pet\}}~~~~~~~$\Longrightarrow$ $\mathbf x_\text{Topic, hot} = [...,0,1,0,...,0,1,0...]^\top$.

\noindent Note that in most industrial  systems, the vocabulary size (especially that of user / author / video ids) can easily scale to hundreds of millions. Thus, a common strategy is to transform the extreme high dimensional hot codes to low dimensional embeddings,
\begin{equation}
\mathbf x_\text{A,emb} = E_A \mathbf x_\text{A,hot},
\end{equation}
where $E_A\in \mathbb R^{d_A \times v_A}$ is the embedding dictionary of $A$, and $d_A$ is the embedding dimension. In our system, we set the embedding dimension to $64$ for id features with large vocabulary, and to $8$ for others, such as video topic, video played timestamp.

In all the upper layers, we take the embedding vectors as input and thus omit the subscript ``$\text{emb}$'' for conciseness.
\subsubsection{The Deep Networks}
The overall architecture of our CTR prediction is shown in Figure \ref{fig:framework}. 
The upper module, consisting of stacked neural networks and ReLUs, acts as a mixer that learns the interaction between the outputs of
three intermediate modules: 
\begin{itemize}[leftmargin=*]
\item \textbf{TWIN}, the proposed consistency-preserved life-long user behavior modeling module, extracts user interest through two cascading stages of behavior modeling sub-modules:
1). Consistency-Preserved General Search Unit (\textbf{CP-GSU}) which performs a coarse search for 100 most relevant behaviors from tens of thousands of long term historical behaviors; 2). Exact Search Unit (\textbf{ESU}) which adopts an attention mechanism over the 100 finalists of CP-GSU to capture the exact user interest.

Different from conventional algorithms that usually consist of a ``light'' GSU and a ``heavy'' ESU,
our proposed CP-GSU follows the identical relevance evaluation metric as that of ESU,
making the two cascading stages \textit{TWINS}. Consequently, CP-GSU consistently retrieves items that are considered important by ESU, maxmizing the behavior modeling effectiveness.
\item \textbf{Short-term behavior modeling} which extracts user interests from the 50 most recent behaviors. This module focuses on the user's short term interest in the latest few days, and acts as a strong complement to TWIN.
\item \textbf{Others Task Modelings}. Besides behavior modeling, we also concatenate the outputs of various other task modelings, which model the user's gender, age, occupation, location, the video's duration, topic, popularity, quality, and contexts features such as the played date, timestamp, page position, etc. 
\end{itemize}

\subsection{TWIN: TWo-stage Interest Network}\label{sec: TWIN}
We name the proposed algorithm TWIN to highlight that CP-GSU follows the identical relevance evaluation metric as that of ESU. Note that this consistency is nontrivial because:
\begin{itemize}[leftmargin=*]
\item Effective behavior modeling algorithms are usually based on Multi-Head Target Attention (MHTA) \cite{vaswani2017attention}, which precisely captures user interest by emphasising target relevant behaviors. 
Unfortunately, due to the high computational complexity, 
the applicable behavior sequence length  of MHTA is mostly limited to a few hundreds.

\item To exhaustively capture user's long-term interest, CP-GSU should cover user behaviors in the last several months, which could easily reach tens of thousands. This sequence length is far beyond the capacity of conventional MHTA given the strict low latency requirements of online systems.  
\end{itemize}

\noindent This section aims to answer this key question: how to improve the efficiency of MHTA so that we can
extend it from ESU to CP-GSU, or namely from a sequence length of hundreds to a sequence length of at least tens of thousands? 

\subsubsection{Behavior Feature Splits and Linear Projection}
Following the standard notations of the MHTA \cite{vaswani2017attention}, we define the features of a length $L$ behavior sequence $[s_1,s_2,...,s_L]$ as matrix $K$, where each row denotes the features of one behavior. 
In practice, the linear projection of $K$ in the attention score computation of MHTA is the key computational bottleneck that hinders the application of MHTA on extremely-long user behavior sequences. We thus propose the followings to reduce its complexity.

We first split the behavior features matrix $K$ into two parts,
\begin{equation}\label{eq:be_splits}
K \triangleq\left[
\begin{array}{cc}
K_{h} & K_c
\end{array}\right] \in \mathbb R^{L \times (H+C)},
\end{equation}
We define $K_h \in \mathbb R^{L \times H}$ as the \textit{inherent} features of behavior items (e.g. video id,  author, topic, duration) which are 
independent of the specific user / behavior sequence,
and $K_c \in \mathbb R^{L \times C}$ as the user-item cross features (e.g. user click timestamp, user play time, clicked page position, user-video interactions).
This split allows high efficient computation of the following linear projection $K_hW^h$ and $K_cW^c$.

For the inherent features $K_h$, although the dimension $H$ is large (64 for each id feature), the linear projection is actually not costly. The inherent features of a specific item are shared across users / behavior sequences. With essential caching strategies, $K_h W^{h}$ could be efficiently ``calculated'' by a look up and gathering procedure. Details of online deployment will be introduced in Section \ref{sec:deploy}.

For the user-item cross features $K_c$, caching strategies are not applicable because: 
1). Cross features describe the interaction details between a user and a video, thus not shared across users behavior sequences. 
2). Each user watches a video for at most once. Namely, there is no duplicated computation in projecting cross features.
We thus reduce the computational cost by simplifying the linear projection weight. 

  Given $J$ cross features, each with embedding dimension $8$ (since not id features with huge vocabulary size). We have $C = 8J$.
 We simplify the linear projection as follows,
\begin{equation}
K_c W^c \triangleq \left[
\begin{array}{ccc}
K_{c,1} \mathbf w_{1}^c,&...&,K_{c,J}\mathbf w_{J}^c
\end{array}\right]
,
\end{equation}
where $K_{c,j} \in \mathbb R^{L \times 8}$ is a column-wise slice of $K_c$ for the $j$-th cross feature, and $\mathbf w_{j}^c \in \mathbb R^{8}$ is its linear projection weight. Using this simplified projection, we compress each cross feature into one dimension, i.e., $K_c W^c \in \mathbb R^{L\times J}$. Note that
this simplified projection is 
equivalent to restricting $W^c$ to a diagonal block matrix. 

\subsubsection{Complexity Analysis} \label{sec:complexity}
In the conventional MHTA, the time complexity of linear projection of $K$, namely from dimension $L\times(H+C)$ to $L\times d_{out}$, is $O(L \times (H+C) \times d_{out})$. 

While in our MHTA for TWIN, the item inherent features $K_hW^h$ is pre-computed and efficiently gathered in $O(L)$, which is independent of the dimension $H$. And the user-item cross features $K_cW^c$ is 
reduced to low dimensional computation of $O(L\times C)$. 

Since $C \ll H$, and $C \ll d_{out}$,
it is this theoretical acceleration that allows the consistent implementation of MHTA in both CP-GSU and ESU.

\subsubsection{Target Attention in TWIN}
Based on the linear projection of behaviors $K_hW^h$ and $K_cW^c$, we now define the target-behavior relevance metric that is used uniformly in both CP-GSU and ESU.

Without loss of generality, we assume that there has been no interaction between the user and the target item and denote the target item's inherent features as $\mathbf q \in \mathbb R ^{H}$. 
With proper linear projection $W^q$, the relevance score $\boldsymbol \alpha \in \mathbb R^L$ between the target item and the historical behaviors is calculated as:
\begin{equation}\label{equ:alpha}
\boldsymbol{\alpha} = \frac{(K_h W^{h})(\mathbf q^\top W^{q})^\top}{\sqrt d_k}+ (K_c W^c)\boldsymbol{\beta},
\end{equation}
where $d_k$ is the dimension of projected query and key.
This relevance score is calculated by the inner product between the \textit{query}, i.e., the inherent features of the target, and the \textit{keys}, i.e., the inherent features of behaviors. Additionally, the cross features, since compressed to 1 dimension, serve as bias terms. We use
 $\boldsymbol \beta \in \mathbb R^J$ as the learnable parameter for the relative importance of cross features. 

In CP-GSU, this relevance score $\boldsymbol \alpha$ is used to cut-off the $L= 10^4$ long-term historical behaviors to 100 most relevant ones.
And in ESU, we perform a weighted average pooling over the 100 finalists: 
\begin{equation}
\text{Attention}(\mathbf q^\top W^q,K_hW^h,K_cW^c,KW^v)= \text{Softmax}(\boldsymbol \alpha)^\top K W^v,
\end{equation}
where
$W^v$ is a projection matrix. We slightly abuse the notation by setting $L =100$.
This projection $KW^v$ is only performed over 100 behaviors and thus can be conducted efficiently online. We do not need to split $K$ as we did when computing $\boldsymbol \alpha$ for $10^4$ behaviors.

To jointly attend to information from different representation subspaces, we adopt 4 heads in our MHTA. Thus the final output of TWIN is defined as
\begin{equation}
\begin{aligned}
&\text{TWIN} = \text{Concat}(\text{head}_1,...,\text{head}_4)W^o, \\
\text{head}_a &= \text{Attention}(\mathbf q^\top W_a^q,K_hW_a^h,K_cW_a^c,KW_a^v), a\in\{1,...,4\},
\end{aligned}
\end{equation}
$W^o$ is a projection that learns relative importance between heads. 
\subsection{System Deployment}\label{sec:deploy}
\begin{figure}[t]
  \centering
  \hspace{-0.3cm}
  \includegraphics[width=0.46\textwidth]{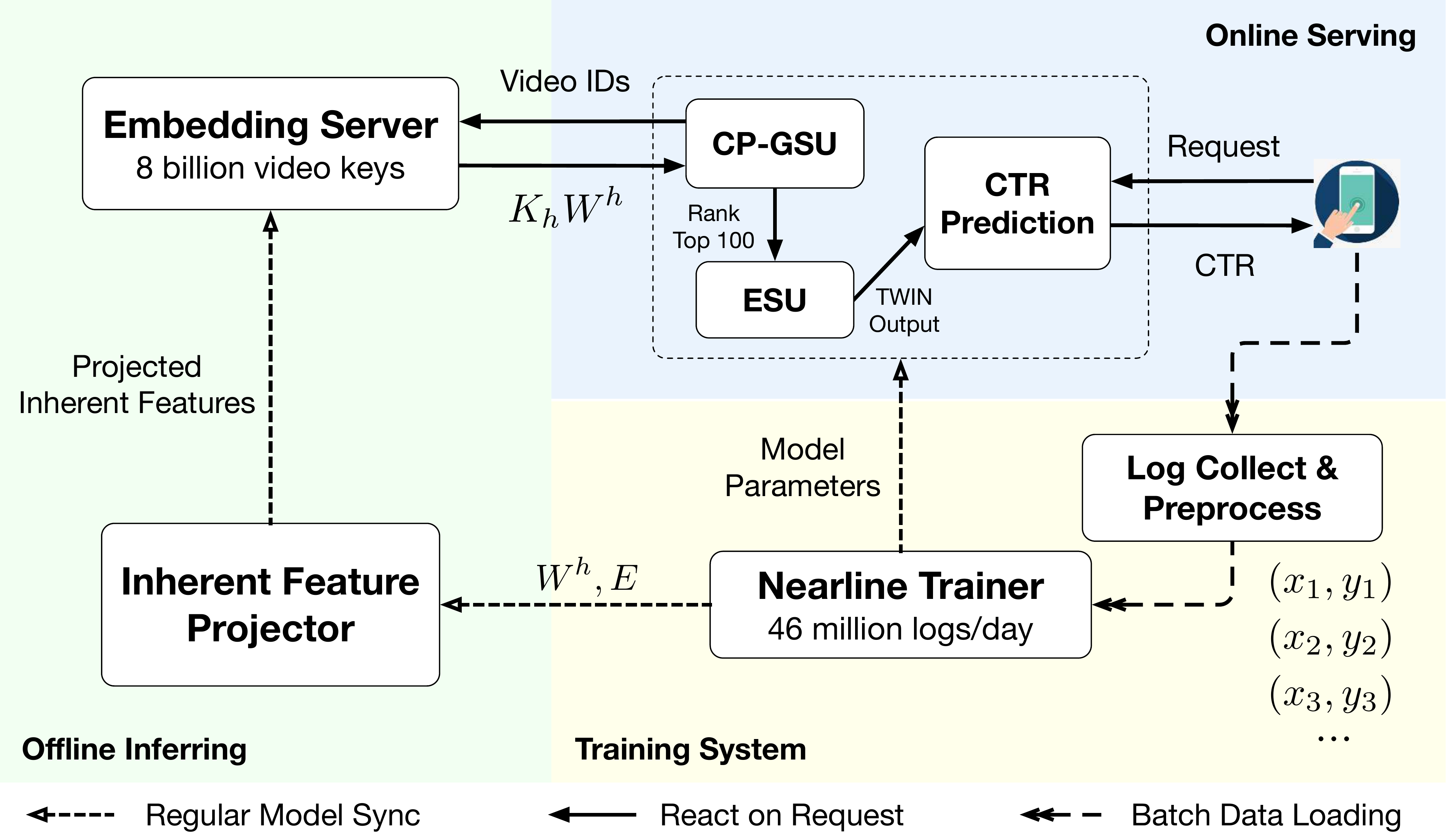}
  \caption{The Deployment of TWIN in Online CTR Prediction System.
  We propose essential precomputing \& caching strategies to reduce the key computational bottleneck, the linear projection of $10^4$ behaviors' inherent features. 
  With proper frequency control, we cut off tailed videos and limit the size of candidate video pool to 8 billion.  
As a result, the \textit{inherent feature projector} can cyclically refresh the linear projection of all candidate videos every 15 minutes, minimizing the accuracy loss from caching. And the \textit{embedding server} which stores the projection of 8 billion candidate videos, can cover 97\% of the online requests and achieve satisfactory effectiveness given limited computational resources.
  }
  \vspace{-0.15cm}
  \label{fig:cbss}
\end{figure}

We deploy TWIN on the ranking system of Kuaishou, serving the main traffic of 346 million daily active users.
In this section, we introduce our hands-on experience in the deployment. Details of our system architecture is shown in Figure \ref{fig:cbss}.

\subsubsection{Training System}
Our TWIN module is trained jointly with the whole CTR prediction model on Kuaishou's large-scale distributed nearline learning system.

Every day, hundreds of millions of users visit Kuaishou, watch and interact with short videos, and leave 46 billion watch and interaction logs per day. Each log is collected, preprocessed in realtime and used for model training in less than 8 minutes. This nearline training system incrementally updates the model parameters using the latest knowledge from user-video interactions that take place in less than 8 minutes ago.

In addition,
our message queue system continuously synchronizes the latest parameter values from the training system to the
offline inferring and online serving systems, once every 5 minutes. This synchronization ensures that the online CTR prediction service is always based on the update-to-date model. 

 \subsubsection{Offline Inferring}\label{sec:delay in cach}
The offline inferring system aims to accelerate the online serving by providing a lookup service.
When receiving lookup keys, a batch of video ids, 
this service returns the lookup values, the corresponding projected inherent features in concatenation, i.e. $K_hW^h_a$ for all heads $a \in \{1,...,4\}$.

Specifically, the offline inferring system consists of two parts. 1). 
\textit{An inherent feature projector}, which cyclically pre-computes the linear projection of inherent features using the latest embeddings and TWIN parameters $W^h_a$ synchronized from the training system. With proper frequency control, this projector can refresh the projected inherent features of an 
8 billion scale candidate video pool every 15 minutes, minimizing the accuracy loss from caching
2). \textit{An embedding server}, which stores results of the inherent feature projector into a key-value structure and provides the aforementioned key lookup service. By cutting off tailed videos, the 8 billion keys can cover 97\% of online request, balancing efficiency and effectiveness. 

\subsubsection{Online Serving}
Once a request is received, the online serving system queries the offline inferring system for the projected inherent features $K_hW^h_a$, and calculates the other parts of Eq \ref{equ:alpha} in realtime for the user-item relevance score $\boldsymbol \alpha$. 
We then select 100 behaviors with the highest attention scores in $\boldsymbol \alpha$ and input the 100 to ESU.
This design reduces the computational bottleneck of TWIN, i.e., linear projection of $K_h$ with $10^4$ rows, by 99.3\% in practice.

Note that ESU with only 100 behaviors is light enough for all calculations to be conducted in realtime using up-to-date parameters synchronized from the training system. As the result, the $K_hW^h$ calculated by ESU is slightly more up-to-date than that in CP-GSU, which further lifts the performance of our TA mechanism.

With accelerated design, TWIN is successfully deployed on the ranking system of Kuaishou, serving the main traffic of 346 million  active users, with a peak request of 30 million videos per second.

\section{EXPERIMENT}
In this section, we detail the offline and online experiments on real industrial data to evaluate our proposed method with the purpose of answering the following four research questions (RQs). 
\begin{itemize}[leftmargin=*,partopsep=0pt,topsep=0pt]
    \setlength{\itemsep}{0pt}
    \setlength{\parsep}{0pt}
    \setlength{\parskip}{0pt}
\item \textbf{RQ1:} How does TWIN perform in the offline evaluations compared to other SOTAs in lifelong user behavior modeling?
\item \textbf{RQ2:} How \textit{consistent} can TWIN achieve, compared to other SOTAs? Or namely, why TWIN works?
\item \textbf{RQ3:} How does the effectiveness of TWIN change as the length of the user behavior sequence grows?
\item \textbf{RQ4:} %What are the effectiveness and efficiency of key model designs in TWIN? 
What are the effects of key components and different implementations in the proposed method?
\item \textbf{RQ5:} How does TWIN perform in real online RS?
\end{itemize}

\vspace{-5px}
\subsection{\textbf{Dataset}}

\begin{table}[t]
\small
\caption{Statistics of the industrial dataset  constructed on the daily collected user logs from Kuaishou.}
\begin{tabular}{llc}
\toprule
Data & Field & Size \\ 
\midrule
\multirow{4}{*}{Daily Log Info} & Users & 345.5 million \\
& Videos & 45.1 million \\
& Samples & 46.2 billion \\
& Average User Actions & 133.7 / day \\
\midrule
\multirow{2}{*}{Historical Behaviors} & Average User Behaviors & 14.5 thousand \\
& Max User Behaviors & 100 thousand \\
\bottomrule
\label{tab:dataset}
\end{tabular}
\vspace{-15px}
\end{table}

To evaluate TWIN in real-world situation for lifelong user behavior modeling, we need a large scale CTR prediction dataset with rich user historical behaviors 
that can ideally scale up to tens of thousands behaviors per user. 
Unfortunately, existing public datasets are either relatively small or lack of sufficient user historical behaviors.
For example, in the widely used Amazon dataset \cite{mcauley2015image, he2017translation, kang2018self} each user has less than ten historical behaviors %, i.e., reviews sorted by timestamp,
 on average. In the Taobao dataset \cite{pi2019practice, zhu2018learning, zhou2018deep, zhou2019deep, pi2020search} the average sequence length of user behaviors is at most 500. We thus collect an industrial dataset from Kuaishou, one of the top short-video sharing platforms in China.

We construct samples from daily user logs, with user's clickthroughs as labels.
As shown in Table \ref{tab:dataset}, the size of daily active users on Kuaishou is around 346 million. 
Every day 45 million short videos are posted and these videos are played 46 billion times in total. On average, each user watches 133.7 short videos per day. %which provides us a testbed to saving user historical behaviors with scale up to thousands. 
To utilize rich behavior information, we collect full user historical behaviors from older logs back to months ago. On average, each user watched 14,500 videos in the past six months, which provides models a testbed with fertile user historical behaviors to learn from. We cutoff the maximum user behavior sequence length to 100,000, which is about the annual total number of views for heavy users.

\vspace{-5px}
\subsection{\textbf{Baselines}}
To demonstrate effectiveness, we compare TWIN with 
the following 
SOTA lifelong user behaviors modeling algorithms. 
\begin{itemize}[leftmargin=*,partopsep=0pt,topsep=0pt]
\setlength{\itemsep}{0pt}
\setlength{\parsep}{0pt}
\setlength{\parskip}{0pt}
\item \textbf{Avg-Pooling}, the average pooling of user lifelong behaviors. %, the most efficient way to summarize user behavior sequences.
\item \textbf{DIN}~\cite{zhou2018deep}, the most widely adopted approach for short-term behavior modeling, which leverages TA for target-specific interests.
\item \textbf{SIM Hard}~\cite{pi2020search}. GSU selects behaviors from the same category with the target item and ESU follows the TA in DIN. In our scenarios, the total number of the video categories is 37.
\item \textbf{ETA } \cite{chen2021end}.
Locality-sensitive hash (LSH) is used to generate a hash signature for the target video and behaviors. 
Then GSU uses Hamming distance as the target-behavior relevance metric.

\item \textbf{SDIM} \cite{cao2022sampling}. GSU selects behaviors with the same hash signature as the target video through multi-round hash collision. In the original paper, ESU linearly aggregates sampled behaviors from multiple rounds to obtain user interests. In our experiments, a more powerful TA is adopted in ESU for fair comparison.

\item \textbf{SIM Cluster}. Since ``category'' requires expensive human annotations and is usually unreliable in short video scenarios, we implement SIM Cluster as an improved variant of SIM Hard.
We group videos into 1,000 clusters based on pre-trained embeddings. GSU retrieves behaviors from the same cluster as the target item. %When the number of behaviors in the same cluster is not enough, it obtains supplements from the most relevant 100 clusters.

\item \textbf{SIM Cluster+} is a refinement of SIM Cluster, where the number of clusters is expanded from 1,000 to 10,000. %, and the number of the nearest neighbor clusters is expanded from 100 to 1,000.
\item \textbf{SIM Soft}~\cite{pi2020search}. GSU uses inner product score of videos' pre-trained embeddings to retrieve relevant behaviors. Inner product is a more refined retrieval method than Hamming distance and hash collision, but with higher computational cost.

\end{itemize}
 \noindent
In summary, ETA and SDIM adopt end-to-end training methods, but use rough retrieval methods to avoid high-complexity calculations. SIM Cluster (+), and SIM Soft use refined retrieval methods, but the compromise is that they have to use pre-trained embeddings and generate offline inverted index in advance.
Note that SIM Soft has not yet been defeated by the follow-up work ETA and SDIM.
We do not compare with UBR4CTR \cite{qin2020user,qin2023learning} because its iterative training is not suitable for our streaming scenario. Furthermore, UBR4CTR is confirmed performing worse than SIM Hard and ETA \cite{chen2021end}.

\vspace{-5px}
\subsection{\textbf{Experimental Setting}}
We use samples in 23 consecutive hours of a day as training data and ones in the following hour for test. We evaluate all algorithms in 5 consecutive days and report the averaged performance over days.
For offline evaluation, we use two widely adopted metrics: AUC and GAUC. AUC signifies the probability that a positive sample's score is higher than that of a negative one, reflecting a model's ranking ability. GAUC performs a weighted average over all user's AUC, and the weights are set to the number of samples of this user. GAUC eliminates the bias among users and evaluates model performance at a finer and fair granularity.

For fair comparison,
in all algorithms, we use identical network structures
including the embedding layers, the upper deep networks, the short-term behavior modeling and other task modelings, except the long-term behavior modeling module.
For two-stage models, we use the recent 10,000 behaviors as input of GSU and retrieve 100 behaviors for TA in ESU.
For DIN, we use the most recent 100 behaviors due to its bottleneck in processing long sequences.
Though CP-GSU of TWIN uses four heads in the attention score computation, we recursively traverses the top ranked items by the four heads until it collects 100 unique behaviors.
For all models, the embedding layer uses the AdaGrad optimizer and the learning rate is 0.05. DNN parameters are updated by Adam with learning rate 5.0e-06. The batch size is set as 8192.

\begin{table}[]
\normalsize
\caption{Offline comparison with SOTAs (RQ1).
We report the \textit{mean} and standard deviation (\textit{std}) over 5 consecutive days. The best and second-best results are highlighted in bold and underlined respectively . Note that the improvement of TWIN over the best compared model in AUC is $0.29\%$ and that in GAUC is $0.51\%$. These improvements are much larger than $0.05\%$, a value enough to bring online benefits.}
\begin{tabular}{lcc}
\toprule
Method & AUC (\textit{mean} \footnotesize{$\pm$ \textit{std}} ) $\uparrow$  & GAUC (\textit{mean} \footnotesize{$\pm$ \textit{std}} ) $\uparrow$  \\ 
\midrule
Avg-Pooling  & 0.7855 \footnotesize{$\pm$ 0.00023} & 0.7168 \footnotesize{$\pm$ 0.00019} \\
DIN  & 0.7873 \footnotesize{$\pm$ 0.00014} & 0.7191 \footnotesize{$\pm$ 0.00012} \\
\hline
SIM Hard  & 0.7901 \footnotesize{$\pm$ 0.00016} & 0.7224 \footnotesize{$\pm$ 0.00021} \\
ETA   & 0.7910 \footnotesize{$\pm$ 0.00004} & 0.7243 \footnotesize{$\pm$ 0.00011} \\
SIM Cluster  & 0.7915 \footnotesize{$\pm$ 0.00017} & 0.7253 \footnotesize{$\pm$ 0.00018} \\
SDIM  & 0.7919 \footnotesize{$\pm$ 0.00009} & 0.7267 \footnotesize{$\pm$ 0.00006} \\
SIM Cluster+  & 0.7927 \footnotesize{$\pm$ 0.00009} & 0.7275 \footnotesize{$\pm$ 0.00011} \\
SIM Soft  & \underline{0.7939 \footnotesize{$\pm$ 0.00014}} & \underline{0.7299 \footnotesize{$\pm$ 0.00013}} \\
\midrule
TWIN  & \textbf{0.7962 \footnotesize{$\pm$ 0.00008}} & \textbf{0.7336 \footnotesize{$\pm$ 0.00011}} \\
Improvement & +0.29\% & +0.51\% \\
\bottomrule
\label{tab:overall}
\end{tabular}
 \vspace{-15px}
\end{table}

\vspace{-5px}
\subsection{Overall Performance (RQ1)}
Table~\ref{tab:overall} shows the performance of all models. Note that due to the large amount of users and samples in our dataset, an improvement of $0.05\%$ in AUC and GAUC in the offline evaluation is significant enough to bring online gains for the business.

\textbf{First, TWIN significantly outperforms all baselines, especially two-stage SOTAs with inconsistent GSU.} This validates the key advantage of TWIN in life-long behavior modeling, i.e., the \textit{powerful} and \textit{consistent} TA in CP-GSU.
Specifically, CP-GSU precisely retrieves behaviors that are considered highly relevant by ESU, saving ESU's precious computational resources for the most important user information.
While in others, the ineffective and inconsistent GSU may miss important behaviors and introduce noisy ones, degrading TA performance.
In addition, the gain from Avg-pooling to DIN shows the ability of TA in retrieving effective information. And the gain from DIN to other two-stage SOTAs verifies the necessity of modeling long behaviors. These two together support our motivation: extending TA to long sequences.

\textbf{Second, only end-to-end training is not enough.}
We observe that TWIN apparently outperforms ETA and SDIM, 
two strong baselines whose embeddings in GSU are also trained in end-to-end manner.
Specially, ETA uses LSH \& Hamming distance and SDIM uses multiround hash collision. Both GSU strategies are less precise compared to TA and inconsistent with
the target-behavior relevance metric used in their ESUs.
While CP-GSU in TWIN is not only trained end-to-end, but also consistent with TA in ESU.
This shows that a precise relevance metric is crucial to GSU, validating our advantage over the existing end-to-end algorithms.

 \textbf{Third, modeling lifelong behaviors at a finer granularity is effective.} 
We compare the variants of SIM: SIM Hard with 37 categories, SIM Cluster(+) with 1,000 
 / 10,000 clusters and SIM Soft where the target-behavior relevance score is calculated individually for each behavior. We observe consistent performance improvement as every finer granularity retrieval method is used in GSU.
This is because GSU retrieves behaviors more accurately when it can capture the relevance score between videos more finely.
From this perspective, we further 
attribute our advantage over SIM Soft to the fact that TWIN adopts more accurate relevance metric.

\begin{figure}[t]
  \centering
\hspace{-0.3cm}
    \includegraphics[width=0.38\textwidth]{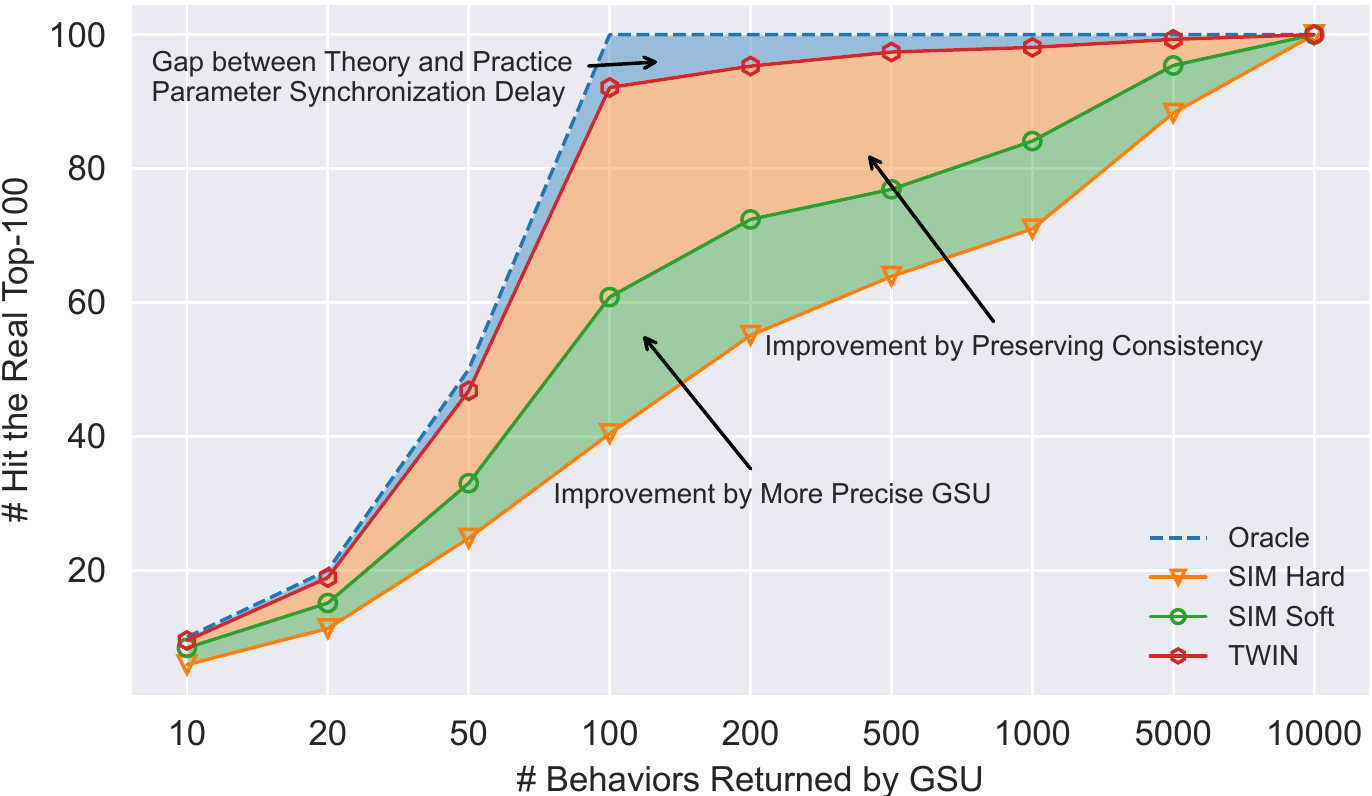}
  \vspace{-5px}
  \caption{Consistency between GSU \& ESU (RQ2).
Oracle uses ESU's relevance metric on $10^4$ behaviors to get ``the real top-100''. Among the 100 return by GSU of SIM Hard, only 40 hits. TWIN lifts the hit number to 94. Theoretically, TWIN could achieve 100\% hit rate since we adopt consistent GSU. However, our practical performance is limited by the deployment constrain, the 15 minutes refresh delay in caching. 
  }
  \vspace{-12px}
\label{fig:consis}
\end{figure}

\subsection{Consistency Analysis (RQ2)}\label{sec:rq2}
As claimed, our superior behavior modeling ability comes from the consistent relevance metrics in CP-GSU and ESU. 
But how consistent we really achieve (Figure~\ref{fig:consis})?

For each well-trained two-stage model, we reuse the parameters of its ESU as its Oracle
to retrieve ``the real top-100'' from 10,000 behaviors.
In other words, these real top-100 are the ground-truth that ESU considers to be really important.
We then traverse the GSU output size of all compared algorithms
from 10 to $10^4$ to examine how many outputs hit the real top-100. 
Note that each compared algorithm has its own Oracle and top-100. But we only plot one Oracle curve since all Oracles perfectly hit ground truth.

SIM Soft achieves an improvement in retrieval consistency benefiting from a more precise retrieval strategy in GSU.
Further, TWIN achieves 94 hits when returning 100 behaviors, which validates our advantage in preserving the consistency between two stages. 
Note that this is the most noteworthy value, since 100 is the upper bound for ESU input, given the constraints of inference time and computational complexity of TA. 
We did not reach our theoretical 100\% consistency in practice due to refresh delay in caching, as described in Section \ref{sec:delay in cach}.
% In addition, we show some case studies of retrieval consistency in the Appendix.

Based on the above results, we speculate that CP-GSU has a stronger ability to match user interests with the target video than conventional GSUs. This is attributed to the consistency. Via sharing the same structure and parameters with ESU, CP-GSU is able to accurately judge and retrieve videos with similar content. %through collaborative filtering signals among similar users. 
In addition, as the CP-GSU parameters are updated in real time, the model can capture the user's dynamically evolving interests.

\vspace{-3px}
\subsection{Effects of Behavior Length (RQ3)}
\begin{figure}[t]
  \centering
\hspace{-0.3cm}
  \subfigure{\label{fig:length-auc}
    \includegraphics[width=0.23\textwidth]{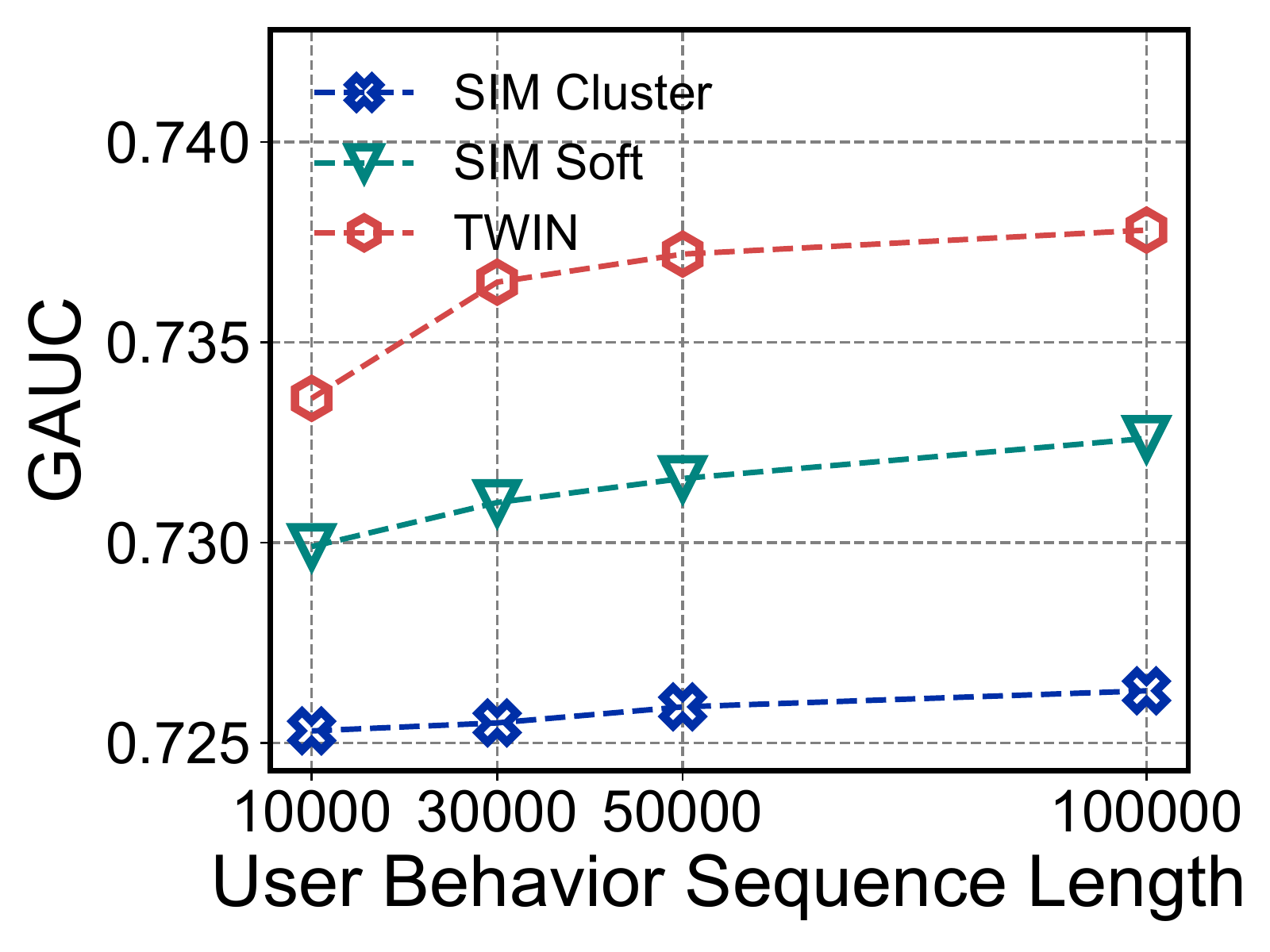}}
    \subfigure{\label{fig:length-wuauc}
    \includegraphics[width=0.23\textwidth]{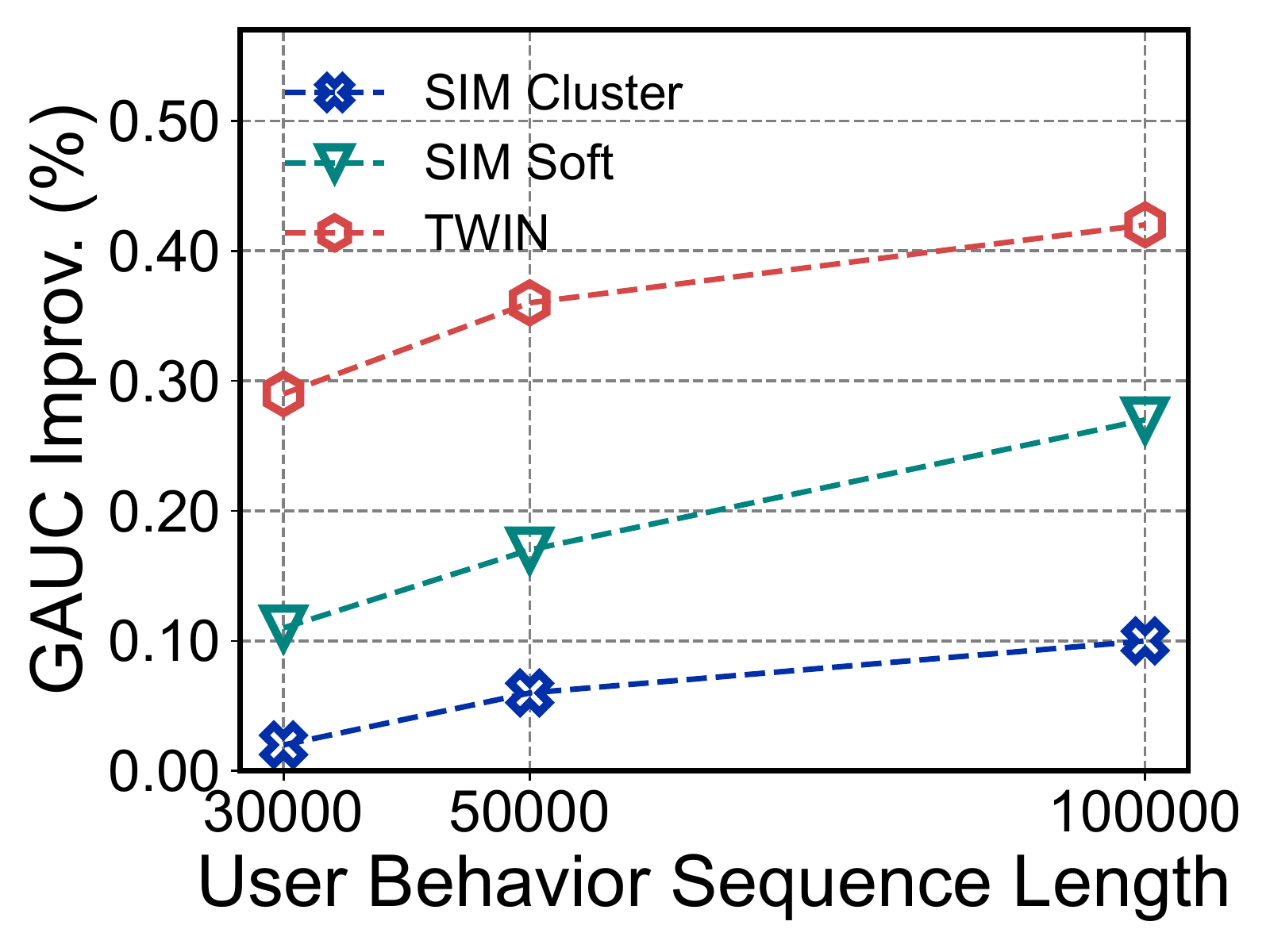}}
\vspace{-0.2cm}
  \caption{
  Performance when GSU Takes Different Behavior Sequence Lengths (RQ3).
  Left: GAUC values.
  Right: Relative GAUC improvements over length = 10,000.
  As the sequence length increases, all model's performance improves and the performance gap between TWIN and baselines increases.}
  \vspace{-0.3cm}
\label{fig:length}
\end{figure}
We aim to test the effectiveness of TWIN under different behavior sequence lengths and to further tap the potential of TWIN. 
Note that only input sequence lengths of GSUs
are changed and the output length is kept 100. Results are shown in Figure~\ref{fig:length}.

We observe that 1). TWIN performs the best consistently, and 2) the performance gap between TWIN and other methods becomes larger as the sequence length increases. %The performance growth of SIM Cluster and SIM Soft is relatively slow compared to that of TWIN. 
This indicates that TWIN has a better effectiveness in modeling extremely long sequences.

\vspace{-3px}
\subsection{Ablation Study (RQ4)}\label{sec:ablation}

\begin{figure}[t]
  \centering
  \hspace{-0.3cm}
  \subfigure{\label{fig:length-auc}
    \includegraphics[width=0.223\textwidth]{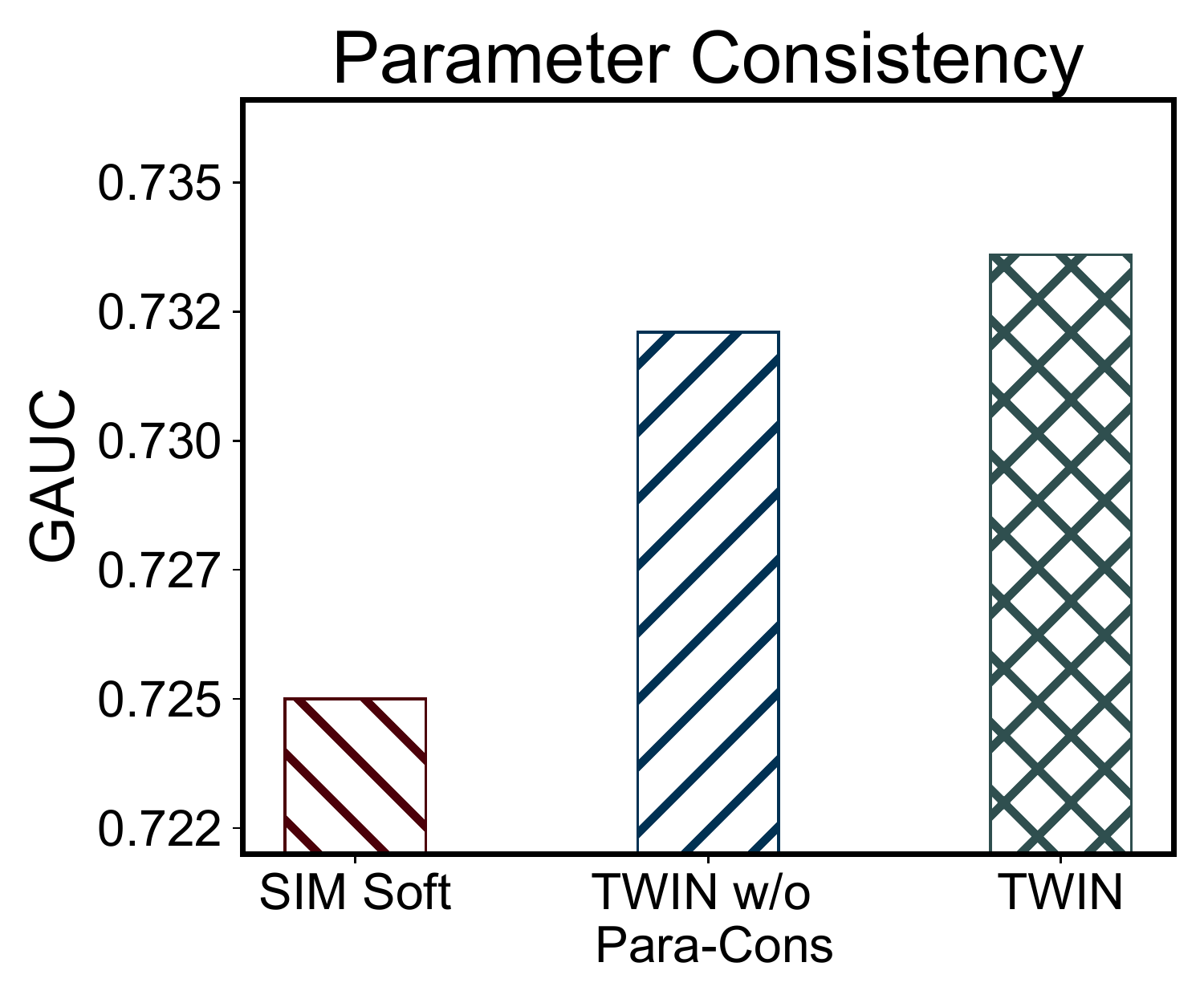}}
   \hspace{-0.3cm}
    \subfigure{\label{fig:length-wuauc}
    \includegraphics[width=0.26\textwidth]{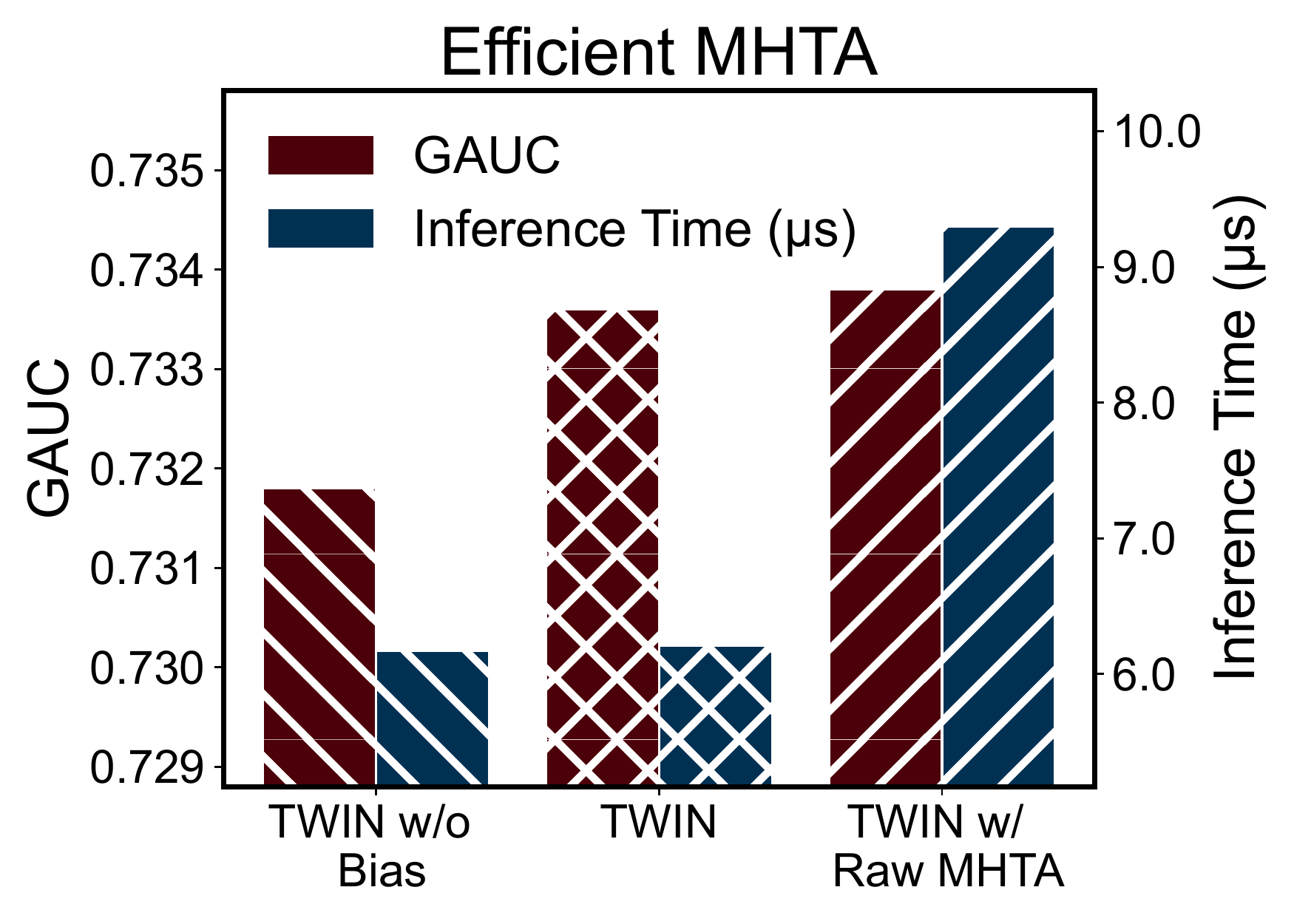}}
  \vspace{-0.2cm}
  \caption{Effects of key Components in TWIN (RQ4). 
  Left: \textit{TWIN w/o Para-Cons} (consistent network structure + inconsistent parameters) outperforms SIM Soft (both inconsistent), but performs worse than TWIN (both consistent).
  Right: directly removing the user-item cross features (\textit{TWIN w/o Bias}) saves little computation but leads to a significant drop in GAUC compared to TWIN. 
  Compared to TWIN with the original MHTA (\textit{TWIN w/ Raw MHTA}), compressing the item inherent features to biases for efficiency (TWIN) hardly harms the performance but greatly speeds up inference time.}
  \vspace{-5px}
\label{fig:ablation}
\end{figure}
We conduct ablation studies by applying different operations to TWIN to evaluate the contribution of our key model designs: 1) the consistency between two stages, and 2) efficient MHTA.

TWIN preserves the consistency between two stages in two folds: the network structure and parameters. To study the benefits brought by each, we implement a variant named \textit{TWIN w/o Para-Con}, that doesn't preserve the parameter consistency. 
Specially, we first train an auxiliary model \textit{TWIN-aux} which uses the identical network structure and training data with TWIN but is trained separately. We then synchronize the GSU parameters from \textit{TWIN-aux} to \textit{TWIN w/o Para-Con}. This is to ensure that \textit{TWIN w/o Para-Con} is still updated in real time and that the gap between TWIN and \textit{TWIN w/o Para-Con} are all caused by parameter inconsistency.

As shown in Figure~\ref{fig:ablation} (left), \textit{TWIN w/o Para-Con} performs significantly better than SIM Soft (inconsistent in both structure and parameters) but slightly worse than TWIN.
It indicates that both structure consistency and parameter consistency are beneficial, but network structure consistency contributes more.

To compute MHTA efficiently for the industrial deployment, we split user behavior features and compress each user-item cross feature into a one-dimensional bias term. To study the impact of such modification and the benefits of preserving user-item cross features in the attention computation, we implement two variants and compare their performances and inference time: 
TWIN with the original MHTA where a straightforward linearly projection $KW$ is used and no feature split is conducted, and TWIN without using the user-item cross features in MHTA, which are respectively abbreviated as \textit{TWIN w/ Raw MHTA} and \textit{TWIN w/o Bias}. 

As shown in Figure~\ref{fig:ablation} (right), TWIN beats \textit{TWIN w/o Bias} significantly and performs almost the same as \textit{TWIN w/ Raw MHTA}, validating that our proposed modification to MHTA hardly compromises the performance.
Regarding to the computational cost, as caching is inapplicable for \textit{TWIN w/ Raw MHTA} when the user-item cross features are used in the linear projection of $K$ (details in Section~\ref{sec:complexity}), the inference time of \textit{TWIN w/ Raw MHTA} significantly increases. In contrast, removing the user-item cross features (TWIN w/o Bias) does not save much computation, yet harms performance.

\begin{table}[t]
\small
\centering
\caption{The relative Watch Time improvement of TWIN in online A/B test compared with SIM Hard and SIM Soft (RQ5). 
In Kuaishou's 
scenario, 0.1\% increase is a significant improvement that brings great business effectiveness.
}
\label{tab::abtest}
\setlength{\tabcolsep}{0.9mm}{\begin{tabular}{lcccc}
\toprule
\textbf{Scenarios}
& \textbf{Featured-Video Tab}
& \textbf{Discovery Tab}
& \textbf{Slide Tab} \\ 
\midrule
\textbf{v.s. SIM Hard}
& +4.893\% & +3.712\% & +6.249\% \\
\textbf{v.s. SIM Soft}
& +2.778\% & +1.374\% & +2.705\% \\
\bottomrule            
\end{tabular}
}
\end{table}

\subsection{Online Result (RQ5)}

To evaluate the online performance of TWIN, we conduct strict online A/B tests on Kuaishou's short-video recommendation platform. Table~\ref{tab::abtest} compares the performance of TWIN with SIM Hard and SIM Soft under three representative business scenarios in Kuaishou (\textbf{Featured-Video} Tab, \textbf{Discovery} Tab, and \textbf{Slide} Tab). 
Different from 
e-commerce, where the commonly used online evaluation metrics are 
CTR and GMV, the short-video recommendation scenarios 
usually use Watch Time, which measures the total amount of time users spend on viewing the videos. As shown, TWIN remarkably outperforms SIM Hard and SIM Soft in all scenarios. As 0.1\% increase in Watch Time is considered as an effective improvement in Kuaishou, TWIN achieves significant business gain. %Currently, TWIN is deployed on Kuaishou App in more than 10 scenarios and serves over 350 million users.

\section{Conclusion}
To address the inconsistency issue of conventional life-long behavior modeling algorithms,
we proposed a consistency-preserved TWo-stage Interest Model, which successfully extended the effective but computational expensive MHTA from ESU to CP-GSU, or namely from a sequence length of $100$ to length $10^4 - 10^5$.
Specifically, we designed novel MHTA mechanism as well as highly efficient infrastructure, including behavior features splits \& compression, pre-computing \& caching, online training \& parameter synchronization. We accelerated the computational bottleneck by 99.3\%, which contributed to the successful deployment of TWIN on Kuaishou, serving the main traffic of hundreds of millions of active users. 
The consistency between two stages, together with the effective TA-based relevance metric in CP-GSU, maximized the retrieval effectiveness of behavior modeling and significantly lifted the performance of CTR prediction.
To the best of our knowledge, TWIN is the first to achieve consistency in the two-stage life-long behavior modeling problem.

\clearpage

\bibliographystyle{ACM-Reference-Format}
\balance
\bibliography{sample-base}

\end{document}